\documentclass[12pt]{article}
\usepackage{amssymb}
\input epsf
\makeatletter
\def\numberbysection{\@addtoreset{equation}{section}
 	\def\theequation{\thesection.\arabic{equation}}}
\makeatother

\numberbysection

\newcommand{\be}{\begin{eqnarray}}
\newcommand{\ee}{\end{eqnarray}}
\newcommand{\non}{\nonumber}

\newcommand{\id}{\mathbb{I}}
\newcommand{\rr}{\mathfrak{r}}

\newcommand{\Q}{\ensuremath{\mathsf{Q}}}
\newcommand{\R}{\ensuremath{\mathsf{R}}}
\newcommand{\SSS}{\ensuremath{\mathsf{S}}}

\newcommand{\Y}{\ensuremath{\mathsf{Y}}}
\newcommand{\Z}{\ensuremath{\mathsf{Z}}}

\newcommand{\LLL}{\ensuremath{\Lambda}}

\begin{document}

\begin{titlepage}
\strut\hfill UMTG--225
\vspace{.5in}
\begin{center}

\LARGE The scaling supersymmetric Yang-Lee model with boundary\\[1.0in]
\large Changrim Ahn\footnote{Department of Physics, Ewha Womans 
University, Seoul 120-750, South Korea} and 
Rafael I. Nepomechie\footnote{Physics Department, 
P.O. Box 248046, University of Miami, Coral Gables, FL 33124 USA}\\

\end{center}

\vspace{.5in}

\begin{abstract}
We define the scaling supersymmetric Yang-Lee model with boundary as
the $(1 \,, 3)$ perturbation of the superconformal minimal model
${\cal SM}(2/8)$ (or equivalently, the $(1 \,, 5)$ perturbation of the
conformal minimal model ${\cal M}(3/8)$) with a certain conformal
boundary condition.  We propose the corresponding boundary $S$ matrix,
which is not diagonal for general values of the boundary parameter. 
We argue that the model has an integral of motion corresponding to an
unbroken supersymmetry, and that the proposed $S$ matrix commutes with
a similar quantity.  We also show by means of a boundary TBA analysis
that the proposed boundary $S$ matrix is consistent with massless flow
away from the ultraviolet conformal boundary condition.
\end{abstract}

\end{titlepage}

\setcounter{footnote}{0}

\section{Introduction}\label{sec:intro}

A $1+1$-dimensional massive integrable quantum field theory without
boundary (i.e., on the full line $x \in (-\infty \,, \infty)$) is
characterized by its factorizable bulk scattering ($S$) matrix
\cite{ZZ1}.  It can also be characterized as a perturbation 
\cite{Za1} of a bulk conformal field theory (CFT) \cite{BPZ}. For 
example, a perturbed minimal model is the renormalization group
infrared trivial fixed point of the action 
\be
A = A_{CFT} + \lambda \int_{-\infty}^{\infty} dy 
\int_{-\infty}^{\infty} dx\  
\Phi_{(\Delta \,, \Delta)}(x \,, y) \,,
\label{bulkaction}
\ee
where $A_{CFT}$ is the action of a $c<1$ minimal model ${\cal
M}(p/q)$, $\Phi_{(\Delta \,, \Delta)}$ is a spinless degenerate
primary field with (right, left) conformal dimensions $(\Delta \,,
\Delta)$ which is relevant ($\Delta < 1$) and ``integrable'', and
$\lambda$ is a parameter of dimension $[\mbox{length}]^{2 \Delta -2}$. 
One link between these two descriptions is provided by the
thermodynamic Bethe Ansatz (TBA), by means of which the central charge
of the CFT can be computed from the $S$ matrix \cite{AlZa1},
\cite{AlZa2}.  The integer-spin and fractional-spin \cite{Za2},
\cite{BL} integrals of motion of an integrable field theory are
manifested in both its $S$ matrix and perturbed CFT descriptions. 
These features of integrable field theory are by now relatively well
understood, due to the great number of examples which have been worked
out in detail.  (See, e.g., \cite{Mu} and references therein.)

For an integrable field theory with boundary (say, on the half-line $x
\in (-\infty \,, 0]$) , the above framework has a nontrivial
generalization \cite{GZ}.  The theory is characterized by a
factorizable {\it boundary} scattering matrix, together with the bulk
$S$ matrix.  It can also be described as a perturbation of a boundary 
CFT. The boundary generalization of (\ref{bulkaction}) is given by
\be
A = A_{CFT + CBC} + \lambda \int_{-\infty}^{\infty} dy 
\int_{-\infty}^{0} dx\ \Phi_{(\Delta \,, \Delta)}(x \,, y)
+  \lambda_{B} \int_{-\infty}^{\infty} dy\  \Phi_{(\Delta)}(y) \,.
\label{boundaryaction}
\ee
The boundary CFT is specified \cite{Ca1} by a conformal boundary
condition (CBC), which for $c<1$ minimal models corresponds to a cell
$(n \,, m)$ of the Kac table.  A CBC is also characterized by the
so-called boundary entropy or ground-state degeneracy ($g$) factor
\cite{AL}, which (roughly speaking) is a measure of the number of bulk
vacua which are compatible with a given CBC. This is well illustrated
in the unitary minimal models \cite{GZ}, \cite{Ch}.  As can
be seen from (\ref{boundaryaction}), the boundary CFT in general has
perturbations by both bulk ($\Phi_{(\Delta \,, \Delta)}$) and boundary
($\Phi_{(\Delta)}$) relevant primary fields.  The boundary parameter
$\lambda_{B}$ has dimension $[\mbox{length}]^{\Delta -1}$.  Note that
the boundary perturbation has the same conformal dimension as the bulk
perturbation, and therefore, presumably it is integrable \cite{GZ}. 
Furthermore, the CBC and the boundary perturbation must be compatible
\cite{Ca1}, \cite{GZ}.  By means of a ``boundary
TBA'' \cite{LMSS} - \cite{AN}, ratios of $g$ factors of
the boundary CFT can be computed from the bulk and boundary $S$
matrices.  (See also \cite{LSS}, \cite{AR}.)  Fractional-spin
integrals of motion should be manifested in both the boundary $S$ matrix and
the perturbed CFT descriptions \cite{MN}.  These features of
integrable field theory with boundary have been studied in relatively
few examples and are less well understood, in comparison to the case
without boundary.

In an effort to provide more such examples, we consider here the 
boundary version of the bulk
scaling supersymmetric Yang-Lee (SYL) model \cite{Sc}-\cite{MS1}. 
This model is arguably the simplest nontrivial supersymmetric quantum
field theory.  Its spectrum consists of one Boson and one Fermion of
equal mass, and the bulk $S$ matrix is factorizable and has $N=1$
supersymmetry.  This model is the supersymmetric generalization of the
scaling Yang-Lee (YL) model \cite{CM},\cite{Sm1},\cite{AlZa1}, which
describes the scaling region near the Yang-Lee singularity of the
two-dimensional Ising model \cite{YL}, \cite{Fi}. The SYL model is 
the first member of an infinite family of integrable models with 
$N=1$ supersymmetry \cite{Sc}.

In particular, we define the boundary SYL model as a perturbed
boundary CFT, and we propose the corresponding boundary $S$ matrix,
which is not diagonal for general values of the boundary parameter. 
We support this picture by identifying a supersymmetry-like integral
of motion, and by studying massless boundary flow using the boundary
TBA. Some related work was done by Moriconi and Schoutens in
\cite{MS2}.  These authors proposed two {\it diagonal} boundary $S$
matrices for the boundary SYL model, without reference to any
specific boundary conditions.  For a special value of the boundary
parameter, our boundary $S$ matrix differs from one of theirs by a
CDD factor.

The outline of this article is as follows.  In Sec.  \ref{sec:YL}, we
briefly review some necessary results about the YL model, and we
clarify a few subtleties of the boundary TBA. In Sec. 
\ref{sec:bulkSYL}, we review some necessary results about the bulk SYL
model.  We also recall the useful observation \cite{Me} that the critical SYL
model can be formulated as either the superconformal minimal model
${\cal SM}(2/8)$ or the conformal minimal model ${\cal M}(3/8)$.  This
is completely analogous to the well-known fact that the tricritical
Ising model can be formulated as either ${\cal SM}(3/5)$ or ${\cal
M}(4/5)$.  One consequence of this fact is that the SYL model can be
regarded, following \cite{Sm2}, \cite{Ta}, as a restriction of the ZMS model
\cite{DB} - \cite{IK}, as we discuss in an appendix.  Sec. 
\ref{sec:boundSYL} is the heart of the paper.  There we first define
the boundary SYL model as a perturbed boundary CFT, and we argue that
it has an integral of motion corresponding to an unbroken
supersymmetry.  
We then propose the boundary $S$ matrix for the boundary
SYL model.  Our approach is to restrict the boundary $S$ matrix of the
boundary supersymmetric sinh-Gordon model \cite{AN}, by imposing the
various boundary bootstrap constraints \cite{GZ}.  We then show that
the proposed boundary $S$ matrix commutes with a supersymmetry-like
charge.  Finally, we perform a boundary TBA analysis, and show that
the proposed boundary $S$ matrix is consistent with massless flow away
from the ultraviolet conformal boundary condition.  In Sec. 
\ref{sec:discuss} we present a brief discussion of our results.

\section{The YL model}\label{sec:YL}

We now briefly recall the basic results of the scaling Yang-Lee model
which we shall need in subsequent sections to formulate the
supersymmetric generalization.  We also clarify a few subtleties of
the boundary TBA.

\subsection{Bulk}

The critical behavior of the Yang-Lee singularity is described
\cite{Ca2} by the minimal model ${\cal M}(2/5)$.  This is a
(nonunitary) CFT with central charge $c=-22/5$.  There are only two
irreducible representations of the Virasoro algebra, and the
corresponding conformal dimensions $\Delta_{(n \,, m)}$ of the primary
fields are organized into a Kac table in Table \ref{figM25}.
\begin{table}[htb] 
  \centering
  \begin{tabular}{|c|c|c|c|}\hline
    0 & $-\frac{1}{5}$ & $-\frac{1}{5}$ & 0 \\
    \hline
   \end{tabular}
  \caption{Kac table for ${\cal M}(2/5)$}
  \label{figM25}
\end{table}

The scaling Yang-Lee model (without boundary) is defined \cite{CM} by
the perturbed action (\ref{bulkaction}), where the CFT is ${\cal
M}(2/5)$, and $\Delta=\Delta_{(1 \,, 3)} = -\frac{1}{5}$.  Arguments
developed by Zamolodchikov \cite{Za1} imply that this model is
integrable.  The spectrum consists of a single particle of mass $m$,
with energy $E= m \cosh \theta$ and momentum $P = m \sinh \theta$,
where $\theta$ is the rapidity.  The two-particle $S$ matrix for
particles with rapidities $\theta_{1}$ and $\theta_{2}$ is given by
\cite{CM}
\be
S_{YL}(\theta) = {\sinh \theta + i \sin(\frac{2 \pi}{3})\over 
\sinh \theta - i \sin(\frac{2 \pi}{3})} \,,
\label{bulkYL}
\ee
where $\theta = \theta_{1} - \theta_{2}$.
This $S$ matrix has a direct (s) channel pole at $\theta = \frac{i 2
\pi}{3}$, since the particle is a bound state of itself. Hence,
the $S$ matrix obeys the bootstrap equation
\be
S_{YL}(\theta + {i \pi\over 3})\ S_{YL}(\theta - {i \pi\over 3}) = 
S_{YL}(\theta) \,.
\label{bootstrapbulkYL}
\ee
The TBA analysis \cite{AlZa1} demonstrates that this $S$ matrix
correctly reproduces the central charge of the unperturbed CFT. The YL
model can be regarded \cite{Sm1} as a restriction of the sine-Gordon
model in which the solitons are projected out and only the first
breather remains.  Indeed, the $S$ matrix (\ref{bulkYL}) coincides
with that of the first sine-Gordon breather \cite{ArKo}, \cite{ZZ1}
with $\gamma = 16\pi/3$.

\subsection{Boundary} \label{sec:YLbound}

Following \cite{DPTW}, \cite{DRTW} , we consider the boundary YL model
which is defined by the perturbed action (\ref{boundaryaction}), where
the CFT is ${\cal M}(2/5)$, the CBC corresponds to the cell $(1 \,,
3)$ of the Kac table, and $\Delta=\Delta_{(1 \,, 3)} = -\frac{1}{5}$. 
The $(1 \,, 3)$ conformal boundary condition and the $(1 \,, 3)$
boundary perturbation are compatible, since the fusion rule
coefficient $N^{(1 \,, 3)}_{(1 \,, 3)\ (1 \,, 3)}$ is nonvanishing. 
The boundary $S$ matrix $\SSS_{YL}(\theta \,; b)$ is given by \cite{DPTW}
\footnote{We make an effort to distinguish boundary quantities from the 
corresponding bulk quantities by using sans serif letters to denote 
the former, and Roman letters to denote the latter.}
\be
\SSS_{YL}(\theta \,; b) =\left(\frac{1}{2}\right) \left(\frac{3}{2}\right) 
\left(\frac{4}{2}\right)^{-1} \left(\frac{1-b}{2}\right)^{-1} 
\left(\frac{1+b}{2}\right) \left(\frac{5-b}{2}\right) 
\left(\frac{5+b}{2}\right)^{-1},
\label{boundaryYL}
\ee 
where
\be 
\left(x \right) \equiv {\sinh({\theta\over 2} +  {i\pi x \over 6})\over
\sinh({\theta\over 2} -  {i\pi x \over 6})} \,,
\label{notation}
\ee
and $b$ is a parameter which is related to $\lambda_{B}$.  This $S$
matrix obeys the boundary bootstrap equation \cite{GZ}
\be
\SSS_{YL}(\theta + {i \pi\over 3}\,; b)\ S_{YL}(2\theta)\ 
\SSS_{YL}(\theta - {i \pi\over 3}\,; b) = \SSS_{YL}(\theta\,; b) \,.
\label{bootstrapboundaryYL}
\ee
This model can be regarded as a restriction of the boundary
sine-Gordon model.  Indeed, the boundary $S$ matrix (\ref{boundaryYL})
coincides with that of the first sine-Gordon breather \cite{Gh} with
$\gamma = 16\pi/3$, and with the parameters $\eta \,, \vartheta$ of 
\cite{GZ} taking the values \cite{DRTW} 
$\eta = \frac{\pi}{4}(b+4) \,, i\vartheta =  \frac{\pi}{4}(b+2)$.

This picture is supported by the boundary TBA, which 
implies that the boundary entropy is given (up to an additive 
constant) by
\be
\ln g = \frac{1}{4\pi}\int_{-\infty}^{\infty}d\theta\ \left[
\kappa_{YL}(\theta\,; b) - \Phi_{YL}(2 \theta) 
- \frac{1}{2}\Phi_{YL}(\theta) \right] L(\theta) \,, 
\label{YLg}
\ee
where 
\be
\Phi_{YL}(\theta) = {1\over i}{\partial\over \partial \theta} 
\ln S_{YL}(\theta) \,, \qquad 
\kappa_{YL}(\theta\,; b) = 
{1\over i}{\partial\over \partial \theta} 
\ln \SSS_{YL}(\theta\,; b) \,,
\ee 
and
\be
\qquad L(\theta) = \ln (1 + e^{-\epsilon(\theta)}) \,. 
\ee 
Moreover, $\epsilon(\theta)$ is the solution of the bulk TBA 
equation \cite{AlZa1}
\be
\epsilon(\theta) = r \cosh \theta 
- {1\over 2 \pi} (\Phi_{YL} * L)(\theta) \,,
\label{YLTBA}
\ee 
where $*$ denotes convolution
\be
\left( f * g \right)(\theta) = \int_{-\infty}^{\infty} 
d\theta'\ f(\theta-\theta')g(\theta') \,,
\ee
and $r=m R$, with $R$ the inverse temperature.
Note that our expression (\ref{YLg}) for the boundary entropy differs
in the third term in the brackets from the one given in Refs. 
\cite{LMSS} and \cite{DRTW}.  This term originates from the exclusion
\cite{FS},\cite{GMN} of the Bethe Ansatz root at zero rapidity.

For simplicity, let us consider the case of massless boundary flow. 
\footnote{The bulk-massive case seems to have several complicated 
issues which remain to be resolved \cite{DRTW}.}
That is, we consider the bulk massless scaling limit 
\be
m = \mu n\,, \qquad \theta = \hat \theta \mp \ln \frac{n}{2} \,, 
\qquad n \rightarrow 0 \,,
\label{limit}
\ee
where $\mu$ and $\hat \theta$ are finite, which implies $E = \mu e^{\pm \hat 
\theta}$, $P = \pm \mu e^{\pm \hat \theta}$. Moreover, we consider
\be
b = -3 - \frac{i 6}{\pi}(\theta_{B} - \ln \frac{n}{2}) \,,
\qquad n \rightarrow 0 \,,
\ee
where the boundary scale $\theta_{B}$ is finite.  For the sign $-$ in
the limit (\ref{limit}), the boundary $S$ matrix reduces to
$S(\hat\theta - \theta_{B})^{-1}$ \cite{DRTW}, and we obtain
\be
\ln g =  - \frac{2}{4\pi}\int_{-\infty}^{\infty}d\hat \theta\ 
\Phi_{YL}(\hat \theta - \theta_{B})\  \hat L(\hat\theta) \,, 
\label{YLg2} 
\ee
where $\hat \epsilon(\hat \theta) \equiv \epsilon(\hat \theta -
\ln \frac{n}{2})$, and $\hat L(\hat\theta) = 
\ln (1 + e^{-\hat \epsilon(\hat \theta)})$. Note the factor of 2 
appearing in (\ref{YLg2}), which accounts for the 
contribution from the sign $+$ in the limit (\ref{limit}).
That is, it can be shown that right-movers and left-movers give equal 
contributions to the boundary entropy. 
In the UV limit $\theta_{B} \rightarrow -\infty$, 
the integrand is nonvanishing for $\hat \theta \rightarrow -\infty$;
similarly, the IR limit $\theta_{B} \rightarrow \infty$ requires
$\hat \theta \rightarrow \infty$. Using the results 
$\hat L(-\infty) =\ln \left({1+\sqrt{5}\over 2}\right)$, $\hat 
L(\infty)=0$ which follow from the TBA Eq. (\ref{YLTBA}), we obtain
\be
\ln {g^{UV}\over g^{IR}} = \ln \left({1+\sqrt{5}\over 2}\right)
\,.
\ee
This is precisely the ratio of $g$ factors corresponding to the
conformal boundary conditions $(1 \,, 3)$ and $(1 \,, 1)$ 
\be
\ln {g_{(1 \,, 3)}\over g_{(1 \,, 1)}} = 
\ln \left({1+\sqrt{5}\over 2}\right)
\,,
\ee
which have been computed \cite{DRTW} from the ${\cal M}(2/5)$ modular
$S$ matrix.  Hence, the boundary $S$ matrix (\ref{boundaryYL}) is
consistent with massless flow away from the UV conformal boundary condition;
namely, from the CBC $(1 \,, 3)$ to the CBC $(1\,, 1)$.  
In Section \ref{sec:SYLboundflow} we shall find a
generalization to the supersymmetric case.

\section{The bulk SYL model}\label{sec:bulkSYL}

We turn now to the supersymmetric generalization of the scaling
Yang-Lee model, which was first defined in \cite{Sc} as a perturbation
of the superconformal minimal model ${\cal SM}(2/8)$.  This
(nonunitary) CFT has central charge $c=-21/4$; and the corresponding
dimensions $\Delta_{(n \,, m)}$ of the primary superconformal fields
are given in Table \ref{figSM28}.  These fields are of Neveu-Schwarz
(NS) or Ramond (R) type if $n-m$ is even or odd, respectively.  We
recall \cite{BPZ} that the superconformal symmetry is generated by the
right and left supercurrents $G(z)$ and $\bar G(\bar z)$ of dimensions
$(\frac{3}{2} \,,0)$ and $(0 \,, \frac{3}{2})$, respectively.  The NS
fields are local with respect to $G(z)$ and $\bar G(\bar z)$, while
the R fields are semi-local with respect to these currents.
\begin{table}[htb] 
  \centering
  \begin{tabular}{|c|c|c|c|c|c|c|}\hline
    0 & $-\frac{3}{32}$ & $-\frac{1}{4}$ & $-\frac{7}{32}$ & 
    $-\frac{1}{4}$ & $-\frac{3}{32}$ & 0 \\
    \hline
   \end{tabular}
  \caption{Kac table for ${\cal SM}(2/8)$}
  \label{figSM28}
\end{table}

The action of the SYL model is given by \cite{Sc}
\be
A = A_{{\cal SM}(2/8)} + \lambda \int_{-\infty}^{\infty} dy 
\int_{-\infty}^{\infty} dx\
G_{-\frac{1}{2}}\bar G_{-\frac{1}{2}}
\Phi_{(\Delta \,, \Delta)}(x \,, y) \,,
\label{bulkSYL}
\ee
where $\Delta=\Delta_{(1 \,, 3)} = -\frac{1}{4}$, and $G_{n}$ 
($\bar G_{n}$) are operators appearing in the operator expansion of the 
supercurrent $G(z)$ ($\bar G(\bar z)$) with 
$\Phi_{(\Delta \,, \Delta)}(z \,, \bar z)$.  An interesting feature of
this model is that it has fractional ($\frac{1}{2}$) spin integrals
of motion.  Indeed, the perturbation preserves supersymmetry, since
\cite{Za2},
\cite{Sc}
\be
\partial_{\bar z} G &=& \partial_{z} \bar \Psi \,, \qquad 
\bar \Psi = \lambda (2 \Delta - 1) \bar G_{-\frac{1}{2}}
\Phi_{(\Delta \,, \Delta)} 
\,, \non \\
\partial_{z} \bar G &=& \partial_{\bar z} \Psi \,, \qquad 
\Psi = \lambda (2 \Delta - 1) G_{-\frac{1}{2}}
\Phi_{(\Delta \,, \Delta)}  \,.
\label{susypcft}
\ee
The corresponding integrals of motion are given by
\be
Q = \int_{-\infty}^{\infty} dx \left[ G(x \,, y) 
+ \bar \Psi(x \,, y) \right] 
\,, \qquad 
\bar Q = \int_{-\infty}^{\infty} dx \left[  \bar G(x \,, y) 
+ \Psi(x \,, y) \right] 
\,. 
\label{bulkSUSYcharges}
\ee 

We now recall the important observation \cite{Me} that there is an
equivalent formulation of the SYL model as a perturbation of the
ordinary minimal model ${\cal M}(3/8)$.  \footnote{As mentioned in the
Introduction, this is completely analogous to the well-known fact that
the tricritical Ising model can be formulated as either ${\cal
SM}(3/5)$ or ${\cal M}(4/5)$.} Indeed, ${\cal M}(3/8)$
also has central charge $c=-21/4$.  The corresponding dimensions of
the primary fields are given in Table \ref{figM38}. Note that these 
dimensions either coincide with those for ${\cal SM}(2/8)$ or else 
correspond to their super-descendants. Indeed, the fields of dimension
$\frac{1}{4}$ and $\frac{3}{2}$ correspond to 
$G_{-\frac{1}{2}} \Phi_{(1 \,, 3)}$ and 
$G_{-\frac{1}{2}} L_{-1} \Phi_{(1 \,, 1)}$ 
respectively; and the field of dimension $\frac{25}{32}$ corresponds 
to $G_{-1} \Phi_{(1 \,, 4)}$.
\begin{table}[htb] 
  \centering
  \begin{tabular}{|c|c|c|c|c|c|c|}\hline
    $\frac{3}{2}$ & $\frac{25}{32}$ & $\frac{1}{4}$ & $-\frac{3}{32}$ & 
    $-\frac{1}{4}$ & $-\frac{7}{32}$ & 0 \\
    \hline
     0 & $-\frac{7}{32}$ & $-\frac{1}{4}$ & $-\frac{3}{32}$ & 
    $\frac{1}{4}$ & $\frac{25}{32}$ & $\frac{3}{2}$ \\
    \hline
   \end{tabular}
  \caption{Kac table for ${\cal M}(3/8)$}
  \label{figM38}
\end{table}

The SYL model can therefore also be formulated by the action
(\ref{bulkaction}), where the CFT is the minimal model ${\cal
M}(3/8)$, and $\Delta=\Delta_{(1 \,, 5)} = \frac{1}{4}$.  This is an
integrable perturbation, since \cite{FLZZ} the $(1 \,, 5)$
perturbation of ${\cal M}(p/q)$ is integrable if $2p < q$.  There is a
corresponding formulation of the conservation laws (\ref{susypcft}),
with the supercurrents $G$ and $\bar G$ replaced by the chiral primary
fields $\Phi_{(2 \,, 1)\,, (1 \,, 1)}$ and $\Phi_{(1 \,, 1)\,, (2 \,,
1)}$ respectively, etc.

The spectrum of the SYL model consists of one Boson and one Fermion of
equal mass $m$.  Following \cite{ZZ1},\cite{GZ}, it is convenient to
introduce the Zamolodchikov operators 
$A_{a}(\theta) = {b(\theta) \choose f(\theta)}$ 
which create the corresponding Boson and Fermion asymptotic particle
states,
\be
|A_{a_{1}}(\theta_{1}) A_{a_{2}}(\theta_{2}) \cdots  
A_{a_{N}}(\theta_{N}) \rangle
= A_{a_{1}}(\theta_{1}) A_{a_{2}}(\theta_{2}) \cdots  
A_{a_{N}}(\theta_{N}) |0 \rangle \,.
\ee
This is an ``in state'' or ``out state'' if the rapidities are 
ordered as $\theta_{1} > \theta_{2} > \cdots  > \theta_{N}$ or 
$\theta_{1} < \theta_{2} < \cdots   < \theta_{N}$, respectively. 

The two-particle $S$ matrix is defined by
\be
A_{a_{1}}(\theta_{1}) A_{a_{2}}(\theta_{2}) = 
S_{a_{1} a_{2}}^{b_{1}b_{2}}(\theta_{1}-\theta_{2})
A_{b_{2}}(\theta_{2}) A_{b_{1}}(\theta_{1}) \,.
\ee 
For the SYL model, the $S$ matrix is given by \cite{Sc}
\be
S(\theta) = S_{YL}(\theta)\ S_{SUSY}(\theta) \,,
\label{bulkSmatrix}
\ee
where $S_{YL}(\theta)$ is given by (\ref{bulkYL}). Moreover,
\be
S_{SUSY}(\theta) = Y(\theta)\ R(\theta) \,,
\label{bulkSUSY}
\ee
where $R(\theta)$ is the $4 \times 4$ matrix \footnote{Our 
conventions are such that if $A$ and $B$ are matrices with matrix 
elements $A_{a_{1} a_{2}}$ and $B_{b_{1}b_{2}}$, then the tensor 
product $C = A \otimes B$ has matrix elements
$C_{a_{1} b_{1}}^{a_{2} b_{2}} = A_{a_{1} a_{2}} B_{b_{1}b_{2}}$.}
\be
R(\theta) = \left( \begin{array}{cccc}
	a_{+}(\theta) &0         &0           &d(\theta)    \\
        0             &b         &c(\theta)   &0            \\
	0             &c(\theta) &b           &0            \\
	d(\theta)     &0         &0           &a_{-}(\theta)
\end{array} \right) \,, 
\label{bulkRmatrix}
\ee 
with
\be
a_{\pm}(\theta) = \pm 1 +  {2 i\sin {\pi\over 3}\over \sinh \theta} 
\,, \qquad
b = 1 \,, \qquad c = {i\sin{\pi\over 3}\over \sinh {\theta\over 2}} 
\,, \qquad
d = {\sin {\pi\over 3}\over \cosh {\theta\over 2}} \,.
\label{Rmatrixelements}
\ee
The scalar factor $Y(\theta)$ is given by
\be
Y(\theta) = {\sinh{\theta\over 2}\over \sinh{\theta\over 2} 
+ i \sin {\pi\over 3}} \exp \left( \int_{0}^{\infty} {dt\over t}\
{\sinh (i t \theta/\pi) \sinh {2t\over 3} \sinh {t \over 3}\over \cosh t 
\cosh^{2}{t\over 2}} \right) \,,
\label{bulkYfactor}
\ee
which we find has the following infinite-product representation:
\be
Y(\theta) &=& {\Gamma \left({1\over 2} + {i \theta\over 2 \pi} \right)
\Gamma \left({1\over 2} - {i \theta\over 2 \pi} \right)\over
\Gamma \left(- {i \theta\over 2 \pi} \right)
\Gamma \left(1 + {i \theta\over 2 \pi} \right)} 
\prod_{k=0}^{\infty} \Big\{
{\Gamma \left({3\over 2} + k - {i \theta\over 2 \pi} \right)^{2}
\Gamma \left(1 + k + {i \theta\over 2 \pi} \right)^{2} \over
\Gamma \left({1\over 2} + k + {i \theta\over 2 \pi} \right)^{2}
\Gamma \left(1 + k - {i \theta\over 2 \pi} \right)^{2}}  \non \\
& \times & {\Gamma \left({2\over 3} + k - {i \theta\over 2 \pi} \right)
\Gamma \left({5\over 6} + k + {i \theta\over 2 \pi} \right)
\Gamma \left({1\over 3} + k - {i \theta\over 2 \pi} \right)
\Gamma \left({7\over 6} + k + {i \theta\over 2 \pi} \right)\over
\Gamma \left({5\over 3} + k + {i \theta\over 2 \pi} \right)
\Gamma \left({5\over 6} + k - {i \theta\over 2 \pi} \right)
\Gamma \left({4\over 3} + k + {i \theta\over 2 \pi} \right)
\Gamma \left({7\over 6} + k - {i \theta\over 2 \pi} \right)} \Big\} \,.
\label{prodrep}
\ee
It is convenient to denote the total scalar factor by $Z(\theta)$
\be
Z(\theta) = S_{YL}(\theta)\ Y(\theta) 
=  {\sinh{\theta\over 2}\over \sinh{\theta\over 2} 
- i \sin {\pi\over 3}} \exp \left( -\int_{0}^{\infty} {dt\over t}\
{\sinh (i t \theta/\pi) \sinh {4 t\over 3} \sinh {t\over 3}\over \cosh t 
\cosh^{2}{t\over 2}} \right) 
\,.
\label{bulkZ}
\ee
Hence, the SYL bulk $S$ matrix is given by
\be
S(\theta) = Z(\theta)\ R(\theta) \,,
\label{bulkS}
\ee
where the matrix $R(\theta)$ is given by Eqs.  (\ref{bulkRmatrix}),
(\ref{Rmatrixelements}).  TBA analysis \cite{Ah1},\cite{MS1} shows
that this $S$ matrix correctly reproduces the central charge of the
unperturbed CFT.

In analogy with the YL model, the SYL model can be regarded as a
restriction of the supersymmetric sine-Gordon (SSG) model in which the
solitons are projected out and only the first breather multiplet
remains.  Indeed, the $S$ matrix is that of the first SSG breather
\cite{SW},\cite{Ah2} with $\alpha=1/3$.  In particular, it coincides
with the expression for the $S$ matrix of the supersymmetric
sinh-Gordon model given in \cite{AN} with $B=-1/3$.

In view of the alternative formulation of SYL as the $(1 \,, 5)$
perturbation of ${\cal M}(3/8)$, the SYL model can also be regarded
as a restriction \cite{Sm2}, \cite{Ta} of the Zhiber-Mikhailov-Shabat model 
\cite{DB}-\cite{IK}. Details of this identification are given 
in Appendix A.

We recall \cite{Za2},\cite{Sc} that the supersymmetry charges are
assumed to act as follows: on one-particle states,
\be
Q A_{a}(\theta) &=& q_{ab}(\theta)\ A_{b}(\theta) \,, \qquad
q(\theta) = \sqrt{m}\ e^{\theta\over 2} \left( \begin{array}{cc}
             0 & e^{{i\pi\over 4}}  \\
             e^{- {i\pi\over 4}} & 0
       \end{array} \right)  \,, \non \\
\bar Q A_{a}(\theta) &=& \bar q_{ab}(\theta)\ A_{b}(\theta) \,, \qquad
\bar q(\theta) = \sqrt{m}\ e^{-{\theta\over 2}} \left( \begin{array}{cc}
             0 & e^{-{i\pi\over 4}}  \\
             e^{{i\pi\over 4}} & 0
       \end{array} \right)  \,;
\label{susyaction1}       
\ee
and on multiparticle states,
\be
\lefteqn{Q |A_{a_{1}}(\theta_{1}) \cdots A_{a_{N}}(\theta_{N}) \rangle
= \sum_{l=1}^{N} \left( \prod_{k=1}^{l-1}(-1)^{F_{a_{k}}} \right)}  \non \\
& & |A_{a_{1}}(\theta_{1}) \cdots A_{a_{l-1}}(\theta_{l-1}) 
\left( Q A_{a_{l}}(\theta_{l}) \right) A_{a_{l+1}}(\theta_{l+1}) 
\cdots A_{a_{N}}(\theta_{N}) \rangle \,, \non \\
\lefteqn{\bar Q |A_{a_{1}}(\theta_{1}) \cdots A_{a_{N}}(\theta_{N}) \rangle
= \sum_{l=1}^{N} \left( \prod_{k=1}^{l-1}(-1)^{F_{a_{k}}} \right)}  \non \\
& & |A_{a_{1}}(\theta_{1}) \cdots A_{a_{l-1}}(\theta_{l-1}) 
\left( \bar Q A_{a_{l}}(\theta_{l}) \right) A_{a_{l+1}}(\theta_{l+1}) 
\cdots A_{a_{N}}(\theta_{N}) \rangle \,,
\label{susyaction2}       
\ee 
where $(-1)^{F}$ is $+1$ for a Boson and $-1$ for a Fermion. These 
charges obey the supersymmetry algebra
\be
Q^{2} = E + P \,, & \bar Q^{2} = E - P \,, &
\left\{ Q \,, \bar Q \right\} = 0 \,,  \non \\
& \left\{ Q \,, (-1)^{F} \right\} = \left\{ \bar Q \,, (-1)^{F} \right\} = 
0 \,. &
\ee 
It can be shown \cite{Sc} that the SYL $S$ matrix commutes with the
supersymmetry charges $Q$ and $\bar Q$, as well as with $(-1)^{F}$.

To conclude this section, we demonstrate that the above $S$ matrix
satisfies the bulk bootstrap equations.  We do this in preparation for
our investigation in Sec. \ref{sec:SYLboundS} of the boundary bootstrap
equations, which will help determine the boundary $S$ matrix. Near 
the direct-channel pole at $\theta = \frac{i2\pi}{3}$, the bulk $S$ 
matrix is given by
\be
S(\theta) \simeq -{i \sqrt{3}\ c^{2} \over \theta - \frac{i2\pi}{3}} 
 \left( \begin{array}{cccc}
	3             &0         &0           &\sqrt{3}    \\
        0             &1         &1           &0            \\
	0             &1         &1           &0            \\
	\sqrt{3}      &0         &0           &1
\end{array} \right) \,, 
\ee
where $c =  \exp \left( -{1\over 2}\int_{0}^{\infty} {dt\over t}\
{\sinh^{2}(2t/3) \sinh (t/3)\over \cosh t 
\cosh^{2}(t/2)} \right)$. Hence, the nonvanishing three-particle 
couplings are given by
\be
f^{b}_{b b} = i c \sqrt{3\sqrt{3}} \,, \qquad 
f^{b}_{f f} = f^{f}_{b f} = f^{f}_{f b} =i c \sqrt[4]{3} \,,
\ee
where $b$ and $f$ denote Boson and Fermion, respectively.
Using the infinite-product representation for the scalar factor $Y(\theta)$ 
(\ref{prodrep}), one can prove the identity
\be
{Y(\theta + \frac{i\pi}{3})\  Y(\theta - \frac{i\pi}{3}) \over
Y(\theta)} = {2\sinh ({\theta\over 2} - {i \pi\over 6})
\cosh ({\theta\over 2} + {i \pi\over 6}) \over \sinh \theta} 
\,.
\ee
Recalling the YL bootstrap relation (\ref{bootstrapbulkYL}), it 
follows that the total scalar factor $Z(\theta)$ (\ref{bulkZ}) satisfies
\be
{Z(\theta + \frac{i\pi}{3})\  Z(\theta - \frac{i\pi}{3}) \over
Z(\theta)} = {2\sinh ({\theta\over 2} - {i \pi\over 6})
\cosh ({\theta\over 2} + {i \pi\over 6}) \over \sinh \theta} 
\,.
\ee
With the help of this identity, it is now straightforward to verify 
the bulk bootstrap equations
\be
f^{c}_{a_{1} a_{2}} S^{b b_{3}}_{c a_{3}}(\theta) =
f^{b}_{c_{1} c_{2}} 
S^{c_{1} b_{3}}_{a_{1} c_{3}}(\theta + \frac{i\pi}{3})\ 
S^{c_{2} c_{3}}_{a_{2} a_{3}}(\theta - \frac{i\pi}{3}) \,.
\ee 

\section{The boundary SYL model}\label{sec:boundSYL}

We now address the main problems of defining the boundary SYL
model and determining its boundary $S$ matrix.  

\subsection{Definition of the model as a perturbed CFT} 
\label{sec:boundSYLdef}

As in the bulk case, we can define the boundary SYL model in either of
two ways.  One way is to define the model as a perturbation of the
superconformal minimal model ${\cal SM}(2/8)$ (cf., Eq. 
(\ref{bulkSYL}))
\be
A &=& A_{{\cal SM}(2/8)+SCBC(1 \,, 3)} + 
\lambda \int_{-\infty}^{\infty} dy \int_{-\infty}^{0} dx\ 
G_{-\frac{1}{2}}\bar G_{-\frac{1}{2}}\Phi_{(\Delta \,, \Delta)}(x \,, y) 
\non  \\
&+&   \lambda_{B} \int_{-\infty}^{\infty} dy\ G_{-\frac{1}{2}} 
\Phi_{(\Delta)}(y) \,, 
\label{boundSYL}
\ee
where $\Delta=\Delta_{(1 \,, 3)} = -\frac{1}{4}$.  Indeed, the
arguments of \cite{GZ} suggest that this boundary perturbation is
integrable.  Following \cite{Ca1}, we observe that for the boundary
CFT, superconformal invariance requires that the stress-energy tensors
and supercurrents obey the boundary conditions
\be
\left( T - \bar T \right)\Big\vert_{x=0} = 0 \,, \qquad 
\left( G - \bar G \right)\Big\vert_{x=0} = 0 \,.
\ee
We assume that for a superconformal minimal model, a superconformal 
boundary condition (SCBC) corresponds to a cell of the Kac table, which
in (\ref{boundSYL}) we take to be $(1 \,, 3)$. (See below.)

Although for the case with boundary the supersymmetry charges 
\be
Q = \int_{-\infty}^{0} dx \left[ G(x \,, y) 
+ \bar \Psi(x \,, y) \right] 
\,, \qquad 
\bar Q = \int_{-\infty}^{0} dx \left[  \bar G(x \,, y) 
+ \Psi(x \,, y) \right] 
\label{boundSUSYcharges}
\ee 
(cf., Eq. (\ref{bulkSUSYcharges})) are not conserved, it is plausible that
some combination of these charges (plus a possible boundary term)
survives.  Indeed, following \cite{GZ}, let us first consider the 
massless case $\lambda=0$, and compute the operator product expansion
$[G(y + ix) - G(y - ix)] G_{-\frac{1}{2}} \Phi_{(\Delta)}(y')$. 
We conclude that the quantity
\be
\Q = \int_{-\infty}^{0} dx\ \left[ G(x \,, y) + \bar G(x \,, y) \right] 
+ \Theta(y) \,,
\ee
with $\Theta(y) \propto \lambda_{B} (1 - 2 \Delta) \Phi_{(\Delta)}(y)$ 
is an integral of motion. It is plausible that, for the general 
massive case $\lambda \ne 0$, this becomes
\be
\Q = Q + \bar Q + \Theta \,,
\label{boundSUSY}
\ee
where $Q$ and $\bar Q$ are given in (\ref{boundSUSYcharges}).

Alternatively, we can define the boundary SYL model as a
perturbation of the minimal model ${\cal M}(3/8)$. That is, we can 
define the model by the action
(\ref{boundaryaction}), where the CFT is ${\cal M}(3/8)$, 
$\Delta=\Delta_{(1 \,, 5)} = \frac{1}{4}$, and the CBC is
either $(1 \,, 3)$, $(1 \,, 4)$, or $(1 \,, 5)$. Indeed, these three
conformal boundary conditions are compatible with the $(1 \,, 5)$ boundary 
perturbation, since the corresponding fusion rule coefficients 
$N^{(1 \,, 5)}_{(1 \,, 3)\ (1 \,, 3)}$,
$N^{(1 \,, 5)}_{(1 \,, 4)\ (1 \,, 4)}$ and
$N^{(1 \,, 5)}_{(1 \,, 5)\ (1 \,, 5)}$
are all nonvanishing, as can be seen from Table \ref{figFR}. 
Presumably, only the CBC $(1 \,, 3)$ preserves superconformal
invariance, since only for this CBC does the corresponding dimension
$\Delta_{(1 \,, 3)} = -\frac{1}{4}$ appear in the ${\cal SM}(2/8)$ Kac
Table \ref{figSM28}. Hence, here we shall consider only the CBC $(1 \,, 
3)$, for which case the corresponding action is presumably equivalent to 
(\ref{boundSYL}).
\begin{table}[htb] 
  \centering
  \begin{tabular}{c c c c c}\hline
    (1, 1, 1) & (2, 2, 3) & (3, 3, 3) & (4, 4, 5) & (5, 5, 5) \\
    (1, 2, 2) & (2, 3, 4) & (3, 3, 5) & (4, 4, 7) \\
    (1, 3, 3) & (2, 4, 5) & (3, 4, 4) & (4, 5, 6) \\
    (1, 4, 4) & (2, 5, 6) & (3, 4, 6) \\
    (1, 5, 5) & (2, 6, 7) & (3, 5, 5) \\
    (1, 6, 6)           & & (3, 5, 7) \\
    (1, 7, 7)           & & (3, 6, 6) \\
    \hline
   \end{tabular}
  \caption{Fusion rule coefficients for ${\cal M}(3/8)$. Here we list
  all the triplets $(i \,, j \,, k)$ with $i \le j \le k$ for which 
  $N^{(1 \,, i)}_{(1 \,, j) (1 \,, k)}$ is nonvanishing, and in fact, 
  equal to 1. Note that $N^{(1 \,, i)}_{(1 \,, j) (1 \,, k)}$ is 
  symmetric under the interchange of any pair of indices $(i \,, j \,, k)$.}
  \label{figFR}
\end{table}

We have obtained the ${\cal M}(3/8)$ fusion rule coefficients given in
Table \ref{figFR} using the corresponding modular $S$ matrix.  Indeed,
we recall (see, e.g., \cite{CIZ}) that for ${\cal M}(p/q)$ the modular
$S$ matrix elements are given by
\be
S_{(r \,, s)\ (r' \,, s')} &=& 2 \sqrt{2\over p q} 
(-1)^{r s' + r' s + 1} \sin {\pi q r r'\over p} 
\sin {\pi p s s'\over q} \,, \non  \\
& & 1 \le r \,, r' \le p - 1 \,, \qquad 1 \le s \,, s' \le q - 1 \,. 
\ee
Setting $r=r'=1$, for ${\cal M}(3/8)$ we obtain the result
\be
S=\left( \begin{array}{ccccccc}
-{1\over 2}\sin{3 \pi\over 8}   &{1\over 2 \sqrt{2}}  
&{1\over 2}\sin{\pi\over 8}   
 &-{1\over 2}    &{1\over 2}\sin{\pi\over 8}  &{1\over 2 \sqrt{2}} 
 &-{1\over 2}\sin{3 \pi\over 8} \\
 {1\over 2 \sqrt{2}} &{1\over 2} &{1\over 2 \sqrt{2}} &0  
 &-{1\over 2 \sqrt{2}} 
 &-{1\over 2} &-{1\over 2 \sqrt{2}} \\
 {1\over 2}\sin{\pi\over 8} &{1\over 2 \sqrt{2}}  
 &{1\over 2}\sin{3 \pi\over 8}
 &{1\over 2}  &{1\over 2}\sin{3 \pi\over 8} &{1\over 2 \sqrt{2}}
 &{1\over 2}\sin{\pi\over 8} \\
 -{1\over 2}  &0  &{1\over 2} &0 &-{1\over 2} &0 &{1\over 2} \\
  {1\over 2}\sin{\pi\over 8} &-{1\over 2 \sqrt{2}}  
  &{1\over 2}\sin{3 \pi\over 8}
 &-{1\over 2}  &{1\over 2}\sin{3 \pi\over 8} &-{1\over 2 \sqrt{2}}
 &{1\over 2}\sin{\pi\over 8} \\
 {1\over 2 \sqrt{2}} &-{1\over 2} &{1\over 2 \sqrt{2}} &0  
 &-{1\over 2 \sqrt{2}} 
 &{1\over 2} &-{1\over 2 \sqrt{2}} \\
-{1\over 2}\sin{3 \pi\over 8}   &-{1\over 2 \sqrt{2}}  
&{1\over 2}\sin{\pi\over 8}   
 &{1\over 2}    &{1\over 2}\sin{\pi\over 8}  &-{1\over 2 \sqrt{2}} 
 &-{1\over 2}\sin{3 \pi\over 8} 
\end{array} \right) \,, 
\label{modS}
\ee
where the matrix element $(s \,, s')$ corresponds to 
$S_{(1 \,, s)\ (1 \,, s')}$. This matrix is real, symmetric, 
and unitary, $S\ S^{\dag} = S^{2}= \id$.
Finally, the Verlinde formula \cite{Ve} implies that the
fusion rule coefficients are given by
\be
N^{(1 \,, i)}_{(1 \,, j) (1 \,, k)}=\sum_{l=1}^{7}
{S_{(1 \,, i)\ (1 \,, l)}\ 
S_{(1 \,, j)\ (1 \,, l)}\ S_{(1 \,, k)\ (1 \,, l)}\over 
S_{(1 \,, 1)\ (1 \,, l)}} \,.
\ee

We close this subsection with the computation of $g$ factors for the 
various conformal boundary conditions, which also relies on the 
modular $S$ matrix. As shown in \cite{Ca1}, \cite{DRTW}, the $g$ 
factor for the CBC $(1 \,, s)$ is given by
\be
g_{(1 \,, s)} = {S_{\Omega\ (1 \,, s)}\over \sqrt{| S_{\Omega\ 0} |}}
\,,
\ee
where $0$ denotes the conformal vacuum (which has the property 
$N^{(1 \,, i)}_{0\ (1\,, j)} = \delta^{i}_{j}$), and $\Omega$ is the 
state of lowest dimension. For ${\cal M}(3/8)$, $0$ is $(1 \,, 1)$ and 
$\Omega$ is $(1 \,, 3)$. In this way, we obtain
\be
g_{(1 \,, 4)} = {1\over \sqrt{2 \sin{\pi\over 8}}} \,, \qquad  
g_{(1 \,, 3)} = g_{(1 \,, 5)} = 
{\sin{3\pi\over 8}\over \sqrt{2 \sin{\pi\over 8}}} \,,  \non  \\
g_{(1 \,, 2)} = g_{(1 \,, 6)} = 
{1\over 2 \sqrt{\sin{\pi\over 8}}} \,, \qquad  
g_{(1 \,, 1)} = g_{(1 \,, 7)} = \sqrt{{1\over 2}\sin{\pi\over 8}} \,.
\label{SYLgfactors}
\ee 
It should also be possible to compute $g$ factors from the ${\cal
SM}(2/8)$ modular $S$ matrix.  However, we do not attempt this 
here. \footnote{It is not clear how to compute the
${\cal SM}(2/8)$ modular $S$ matrix directly from the coset 
$su(2)_{2}\oplus su(2)_{m}/su(2)_{2+m}$ with $m=-4/3$ \cite{MSW}, 
since an additional coset field seems to be required.}

\subsection{Boundary $S$ matrix} \label{sec:SYLboundS}

The boundary $S$ matrix $\SSS(\theta)$ is defined as \cite{GZ}
\be
A_{a}(\theta) B = \SSS_{a}^{b}(\theta)\ A_{b}(-\theta) B \,,
\ee 
where here $B$ is the so-called boundary creation operator.  We now
try to determine $\SSS(\theta)$ for the boundary SYL model
(\ref{boundSYL}).  By analogy with the bulk SYL model, as well as with
the boundary YL model, we expect that the boundary $S$ matrix of the
boundary SYL model should be some reduction of that of the boundary
supersymmetric sine-Gordon model \cite{AK}, or equivalently,
the boundary supersymmetric sinh-Gordon model \cite{AN}. We therefore 
consider
\be
\SSS(\theta)= \SSS_{YL}(\theta \,; b)\ 
\SSS_{SUSY}(\theta \,; \phi) \,,
\label{boundarySmatrix}
\ee
where the scalar factor $\SSS_{YL}(\theta \,; b)$ is given by
(\ref{boundaryYL}), and $\SSS_{SUSY}(\theta \,; \phi)$ is given by
\be
\SSS_{SUSY}(\theta \,; \phi) 
= \Y(\theta \,; \phi)\ 
\R(\theta \,; \phi) \,,
\label{boundarySUSY}
\ee 
where $\R(\theta \,; \phi)$ is the $2 \times 2$ matrix 
\be
\R(\theta \,; \phi) = \left( \begin{array}{cc}
{\cal A}_{+} &{\cal B} \\
{\cal B} & {\cal A}_{-} 
\end{array} \right) \,, 
\label{boundaryRmatrix}
\ee 
with matrix elements
\be
{\cal A}_{\pm} = \cosh{\theta\over 2}\ G_{+}
\pm i \sinh{\theta\over 2}\ G_{-} \,, \qquad 
{\cal B} = - i \sinh \theta  \,,
\label{boundelem1}
\ee 
where
\be
& G_{+} = \rr \left( \sinh\phi  +  
{e^{\phi}\sinh^{2}{\theta\over 2}\over 1 - \sin {\pi\over 3}}
\right) \,, \qquad
G_{-} = \rr \left( \cosh\phi  +  
{e^{\phi}\sinh^{2}{\theta\over 2}\over 1 - \sin {\pi\over 3}}
\right) \,, & \non \\ 
&\rr = \left({2 (1 - \sin {\pi\over 3})\over \sin {\pi\over 3}} 
\right)^{1\over 2} \,. &
\label{boundelem2}
\ee
Moreover, $\Y(\theta \,; \phi)$ is a scalar factor given by
\be
\Y(\theta \,; \phi) = 
\Y_{0}(\theta)\ \Y_{1}(\theta \,; \phi)\ F(\theta \,; \phi)
\,,
\label{boundaryY}
\ee
where
\be
\Y_{0}(\theta) &=& {i\over \sqrt{2} \sinh({\theta\over 2} 
+{i \pi\over 4})} \exp \left( -{1\over 2} \int_{0}^{\infty}
{dt\over t}\
{\sinh (2 i t \theta/\pi) \sinh(2t/3) \sinh (t/3)\over \cosh^{2} t 
\cosh^{2}(t/2)} \right) \,, \non \\ 
\Y_{1}(\theta \,; \phi) &=& {1\over \rr \sinh \phi}
{\sin({\pi\over 12} - {\zeta\over 2})
 \sin({\pi\over 12} + {\zeta\over 2})\over 
 \sin({\pi\over 12} - {\zeta\over 2} - {i \theta\over 2})
 \sin({\pi\over 12} + {\zeta\over 2} - {i \theta\over 2})} \non \\
&\times& \exp{ \left( -2 \int_{0}^{\infty} {dt\over t}\
{\sinh (i t \theta/\pi) \sinh(t/3) \cosh (t \zeta/\pi)\over \sinh t 
\cosh(t/2)} \right)} \,,
\label{boundaryYfactors}
\ee
and $\zeta$ is a function of $\phi$ defined by
\be
\cos \zeta = 1 - e^{-2 \phi}(1 - \sin {\pi\over 3})  \,.
\label{zeta}
\ee
The exponential factors of $\Y_{0}(\theta)$ and $\Y_{1}(\theta \,; 
\phi)$ do not have zeros or poles in the physical strip
$0 \le Im \theta \le {\pi\over 2}$,
provided $|\zeta| < {2\pi\over 3}$.
Finally, $F(\theta \,; \phi)$ is a CDD-like factor obeying
\be
F(\theta \,; \phi)\ F(-\theta \,; \phi) = 1 \,, \qquad 
F({i \pi\over 2} + \theta \,; \phi) 
= F({i \pi\over 2} - \theta \,; \phi) \,,
\label{CDD}
\ee 
which is still to be determined.

The above expression for $\SSS_{SUSY}$ essentially coincides with the
one for the supersymmetric sinh-Gordon model given in \cite{AN} with
$B=-{1\over 3}$, $\varepsilon=+1$, $\varphi = \phi +{i \pi\over 2}$
with $\phi$ real, and $r = -i\rr$.  The only differences lie in the CDD
factor $F(\theta \,; \phi)$ (which is absent from \cite{AN}) and the
factor $\Y_{1}$: the expression given here is an analytic continuation
of the one given in \cite{AN}.  The former does not diverge for
$\theta= \pm {i\pi\over 3}$, which is important for implementing the
boundary bootstrap equations, as we shall see 
below (\ref{SYLboundbootstrap}).

The alert reader will have noticed that, while the boundary SYL action
(\ref{boundSYL}) contains only one boundary parameter (namely,
$\lambda_{B}$), the above boundary $S$ matrix seems to contain two
parameters, namely, $b$ and $\phi$.  The key point to realize is that
these two parameters are {\it not} independent.  By demanding that the
boundary $S$ matrix satisfy the various constraints \cite{GZ} arising
from the existence of boundary and bulk bound states, we shall
determine the relation between $\phi$ and $b$ (\ref{desired}), as well
as the CDD factor $F(\theta \,; \phi)$ (\ref{CDDsolution}).

We begin by considering the constraints due to boundary bound states.
In general \cite{GZ}, let $i v^{\alpha}_{0 a}$ be the position of
a pole of the boundary $S$ matrix 
in the physical strip associated with the excited boundary state
$|\alpha \rangle_{B}$, which can be interpreted as a boundary bound
state of particle $A_{a}$ with the boundary ground 
state $|0\rangle_{B}$. Near this pole, the boundary $S$ matrix 
has the form
\be
\SSS_{a}^{b}(\theta) \simeq {i\over 2}
{g^{\alpha}_{a 0} g^{b 0}_{\alpha}\over \theta - i v^{\alpha}_{0 a}} \,,
\label{boundpole}
\ee 
where $g^{\alpha}_{a 0}$ are boundary-particle couplings.

We assume that (as in the bulk) the SYL boundary $S$ matrix inherits
its pole structure from the YL boundary $S$ matrix (\ref{boundaryYL}). 
Therefore, it has \cite{DTW} two boundary bound state poles,
corresponding to excited boundary states $|1 \rangle_{B}\,, |2
\rangle_{B}$, with \footnote{The
subscript $a$ of $v^{\alpha}_{0 a}$ can be dropped, since YL has only
one type of particle.}
\be
v^{1}_{0} = {\pi(b+1)\over 6} \,, \qquad
v^{2}_{0} = {\pi(b-1)\over 6} \,.
\label{vees}
\ee 
It follows from the condition (\ref{boundpole}) and the form
(\ref{boundaryRmatrix}) of the $S$ matrix that for $\theta = 
iv^{\alpha}_{0}$,
\be 
{\cal A}_{+} \propto (g^{b 0}_{\alpha})^{2} \,, \quad
{\cal A}_{-} \propto (g^{f 0}_{\alpha})^{2} \,, \quad
{\cal B} \propto g^{b 0}_{\alpha} g^{f 0}_{\alpha} \,,
\ee 
where the indices $b$ and $f$ again denote Boson and Fermion,
respectively. Hence, we arrive at the important constraint
\be
{{\cal A}_{+}{\cal A}_{-}\over {\cal B}^{2}}
\Big\vert_{\theta = iv^{\alpha}_{0}} = 1 \,.
\label{constraint}
\ee 
This equation gives a relation between the boundary parameter $\phi$
and $v^{\alpha}_{0}$.  As shown in Appendix B, the relation can be
expressed most succinctly in terms of the parameter $\zeta$ defined in
(\ref{zeta}):
\be
\zeta = v^{\alpha}_{0} \pm {\pi\over 6} \,.
\label{zetarelation}
\ee
The above relation can hold for both poles (\ref{vees})
only if
\be
\zeta = {\pi b\over 6} \,.
\label{desired}
\ee 
Eq. (\ref{desired}) is the desired relation between $\phi$ and $b$.
The restriction $|\zeta| < {2\pi\over 3}$ which we found above implies
$|b| < 4$.

We now consider the constraints due to bulk bound states. In view of 
the direct-channel pole of the SYL bulk $S$ matrix at 
$\theta = {i 2\pi\over 3}$, the following boundary bootstrap relations 
must hold \cite{GZ}
\be
f^{ab}_{d} \SSS^{d}_{c}(\theta) = f^{b_{1}a_{1}}_{c} 
\SSS^{a_{2}}_{a_{1}}(\theta + {i\pi\over 3})\ 
S^{b_{2}a}_{b_{1}a_{2}}(2\theta)\
\SSS^{b}_{b_{2}}(\theta - {i\pi\over 3})\,.
\label{SYLboundbootstrap}
\ee
Using infinite-product representations for the scalar factors 
$\Y_{0}(\theta)$, $\Y_{1}(\theta \,; \phi)$, and $Y(\theta)$, 
one can prove the identities
\be
{\Y_{0}(\theta + {i\pi\over 3}) \Y_{0}(\theta - {i\pi\over 3}) 
Y(2\theta) \over \Y_{0}(\theta)}
&=& {i \sqrt{2} \sinh \theta \sinh({\theta\over 2}-{i\pi\over 4})
\over
\sinh(\theta + {i\pi\over 3}) \cosh(\theta - {i\pi\over 3})}
\,, \non  \\
{\Y_{1}(\theta + {i\pi\over 3}\,; \phi)
\Y_{1}(\theta - {i\pi\over 3}\,; \phi)\over
\Y_{1}(\theta \,; \phi)} &=& {1\over \rr \sinh \phi}
\sin \left({\pi\over 12} - {\zeta\over 2}\right)
\sin \left({\pi\over 12} + {\zeta\over 2}\right) \non  \\
&\times & {2\left( 1 + 2\cos 2 \zeta - 2 \cosh 2 \theta - 4 i \cos 
\zeta \sinh \theta \right)\over \cos 3\zeta + i \sinh 3 \theta}
\,.
\ee
With the help of these identities, together with 
(\ref{bootstrapboundaryYL}), one can show that the SYL
boundary bootstrap relations (\ref{SYLboundbootstrap}) 
are satisfied, provided that the CDD factor obeys
\be
{F(\theta + {i\pi\over 3}\,; \phi) 
F(\theta - {i\pi\over 3}\,; \phi) \over F(\theta \,; \phi)} =
{\cos 3\zeta + i \sinh 3 \theta \over 
\cos 3\zeta - i \sinh 3 \theta} \,.
\label{CDDconstraint}
\ee

In addition to the boundary bootstrap relation, another constraint due
to bulk bound states is stated in \cite{GZ}.  Namely, let $i
u^{c}_{ab}$ be the position of the pole of the bulk $S$ matrix
associated with the direct-channel bound state of $A_{a} A_{b}$ which
can be interpreted as the particle $A_{c}$.  If the particles $A_{a}$
and $A_{b}$ have equal mass, then the boundary $S$ matrix must have a
pole at $\theta= {i \bar u^{c}_{ab}\over 2}$, where 
$\bar u^{c}_{ab} = \pi - u^{c}_{ab}$. Furthermore,
\be
\SSS_{\bar a}^{b}(\theta) \simeq -{i\over 2}
{f^{ab}_{c} g^{c}\over \theta - {i \bar u^{c}_{ab}\over 2}} \,,
\label{bulkpole}
\ee 
where $g^{c}$ describes the coupling of $A_{c}$ to the boundary.
The SYL boundary $S$
matrix indeed has such a pole at $\theta = {i \pi\over 6}$.  It follows
from the condition (\ref{bulkpole}) that for $\theta = {i\pi\over 6}$,
\be
{\cal A}_{+} \propto f^{bb}_{b} g^{b} \,, \qquad
{\cal A}_{-} \propto f^{ff}_{b} g^{b} \,;
\ee 
and hence \cite{MS2}
\be
{{\cal A}_{+} \over {\cal A}_{-}}
\Big\vert_{\theta = {i \pi\over 6}} = {f^{bb}_{b}\over f^{ff}_{b}} = 
\sqrt{3} \,.
\ee 
However, this equation is satisfied for arbitrary values of $\phi$,
and so does not provide any further constraints on the $S$ matrix.

The scalar factor $\Y(\theta \,; \phi)$ (\ref{boundaryY}) should not 
have zeros or poles in the physical strip. In 
view of the relation (\ref{desired}), we see that 
the factor $\Y_{1}(\theta \,; \phi)$ has poles at 
$\theta = i (\pm\zeta - {\pi\over 6}) = i \pi ( \pm b - 1)/6$. The pole at 
$\theta = i \pi ( b - 1)/6$ 
is undesirable, since it is physical for $1 < b < 4$. \footnote{The 
pole at $\theta = i \pi ( - b - 1)/6$ is canceled by a corresponding 
zero in the factor $\left({1 + b\over 2}\right)$ from $\SSS_{YL}$.}
Fortunately, we can arrange for this pole to be canceled by a corresponding 
zero of the CDD factor. Indeed, a solution to the CDD constraint Eqs.
(\ref{CDD}) and (\ref{CDDconstraint}) which has a zero
at $\theta = i \pi ( b - 1)/6$ is given by
\be
F(\theta \,; \phi) = 
\left({1-b\over 2}\right)\left({5+b\over 2}\right) \,,
\label{CDDsolution}
\ee
where we have again used the notation (\ref{notation}).

In short, the boundary $S$ matrix which we propose for the boundary 
SYL model (\ref{boundSYL}) is given by 
Eqs. (\ref{boundarySmatrix}) - (\ref{zeta}), (\ref{desired}), 
(\ref{CDDsolution}). This is one of the 
main results of our paper.
Note that our proposed boundary $S$ matrix depends on a single
independent boundary parameter $b$.  The relation of this parameter to
the boundary parameter $\lambda_{B}$ in the action (\ref{boundSYL}) is
not yet known.

One check on this proposal is provided by supersymmetry.  We have
suggested that the SYL model (\ref{boundSYL}) has the integral of
motion $\Q$ given by (\ref{boundSUSY}).  We now demonstrate that our
proposed boundary $S$ matrix commutes with a similar quantity. 
Indeed, let us assume that the supersymmetry charges $Q$ 
and $\bar Q$ act on states according to (\ref{susyaction1}),
(\ref{susyaction2}). It is straightforward to show that the matrix 
$\R(\theta \,; \phi)$ (\ref{boundaryRmatrix}) commutes with
\be
\Q = Q + \bar Q + \gamma (-1)^{F} \,,
\label{boundSUSY2}
\ee
where here $\gamma = - \sqrt{m} \sqrt{-1+{2\over \sqrt{3}}}
e^{-\phi}$. Note that $\Q$ does not anticommute with $(-1)^{F}$, 
unlike usual supersymmetry charges. The appearance of $(-1)^{F}$ in 
$\Q$ should not be too surprising, since similar topological
charges also appear in the fractional-spin integrals of motion of the
boundary sine-Gordon model \cite{MN}. 
Presumably the operator $\Theta$ in (\ref{boundSUSY}) can
be identified with $\gamma (-1)^{F}$.  Note that $\lambda_{B}=0$ (for
which $\Theta$ vanishes) corresponds to $\phi = \infty$, and hence
$b=0$.  For this value of $b$, the boundary $S$ matrix $\SSS(\theta)$ is 
diagonal. We recall that Moriconi and Schoutens proposed \cite{MS2} two 
diagonal boundary $S$ matrices for the boundary SYL model (although without 
reference to any specific boundary conditions), which they 
designated $R^{[1]}_{(1)}$ and $R^{[1]}_{(2)}$. Our boundary $S$ matrix 
for $b=0$ differs from $R^{[1]}_{(2)}$ by the CDD factor, i.e.,
\be
{\SSS(\theta)\over F(\theta \,; \phi)}\Big\vert_{b=0} = 
R^{[1]}_{(2)}(\theta) \,.
\ee 

\subsection{Boundary TBA and massless boundary flow} \label{sec:SYLboundflow}

We have defined the boundary SYL model in Sec.  \ref{sec:boundSYLdef},
and we have proposed the corresponding boundary $S$ matrix in Sec. 
\ref{sec:SYLboundS}.  We shall now demonstrate that this picture is
supported by the boundary TBA. Our analysis is a generalization of the
one for the boundary YL model, which we briefly reviewed in Sec. 
\ref{sec:YLbound}. For simplicity, we again focus our attention on 
the case of massless boundary flow. 

We begin by determining the massless scaling limit. We set
\be
m = \mu n \,, \quad \theta = \hat \theta - \ln \frac{n}{2} \,, \quad 
{i \pi\over 6}(b + a) = \theta_{B} - \ln \frac{n}{2} \,, 
\quad n \rightarrow 0 \,,
\label{SYLlimit}
\ee
with $\mu$, $\hat \theta$, and $\theta_{B}$ real and finite.  Our objective is
to determine the value(s) of $a$ (also real and finite) for which 
the boundary $S$ matrix, in the above limit, remains finite and unitary.
After some computation, we find that $a=6$; and the resulting 
massless boundary $S$ matrix is given by
\be
\SSS(\theta)= \Z(\hat \theta - \theta_{B})\ 
\R(\hat \theta - \theta_{B}) \,,
\ee
where
\be
\Z(\theta) = {\sinh( {\theta\over 2} - {i \pi\over 12})\over
\sinh( {\theta\over 2} - {i 5\pi\over 12})
\sinh( {\theta\over 2} + {i 5\pi\over 12})} \exp \left(
-\int_{0}^{\infty}{dt\over t}\ {\sinh {t\over 3} 
\sinh \left( t ({i \theta\over \pi} - 1) \right) \over \sinh t \cosh {t\over 2} }
\right) \,,
\label{masslessZfactor}
\ee
and 
\be
\R(\theta) =  \left( \begin{array}{cc}
\sinh({\theta\over 2} + {i \pi\over 4}) &-{i \sqrt[4]{3}\over 2} \\
-{i \sqrt[4]{3}\over 2} & \sinh({\theta\over 2} - {i \pi\over 4})
\end{array} \right) \,.
\ee 
Indeed, $\SSS(\theta)$ satisfies the unitarity condition, since
\be
\Z(\theta)\R(\theta)\ \Z(-\theta)\R(-\theta) = \id \,.
\ee 
 
In order to formulate the TBA equations, we consider $N$ particles
with real rapidities $\theta_{1} \,, \ldots \,, \theta_{N}$ in an
interval of length $L$, with bulk $S$ matrix $S(\theta)$ (\ref{bulkS})
and boundary $S$ matrices $\SSS(\theta \,; b_{\pm})$
(\ref{boundarySmatrix}), where the subscripts $\pm$ here denote the left 
and right boundaries. 
As already discussed, the bulk and boundary $S$ matrices of the SYL model 
essentially coincide with those for the
supersymmetric sinh-Gordon model given in \cite{AN} with $B=-{1\over
3}$, $\varepsilon=+1$, $\varphi = \phi +{i \pi\over 2}$, and $r =
-i\rr$.  Hence, the Bethe Ansatz equations and the transfer matrix
eigenvalues for SYL can be easily obtained from \cite{AN}, to 
which we shall henceforth refer as I. From Eq. (I 4.14)
we obtain the Bethe Ansatz equations for $z_{k}^{+}$
\be
\lefteqn{\prod_{j=1}^{N}
{\tanh({1\over 2}(z_{k}^{+} - \theta_{j}) )\over
\tanh({1\over 2}(z_{k}^{+} - \theta_{j}) + {i \pi\over 3})}
{\tanh({1\over 2}(z_{k}^{+} + \theta_{j}) )\over
\tanh({1\over 2}(z_{k}^{+} + \theta_{j}) + {i \pi\over 3})} = 
{\sinh^{2}({1\over 2}({i5\pi\over 6} + z_{k}^{+}))
\over
\sinh^{2}({1\over 2}({i\pi\over 6} - z_{k}^{+}))}} 
\non \\
&\times&\left[{-e^{-\phi_{-}}\sinh^{2}({i\pi\over 12})
+ e^{\phi_{-}}\sinh^{2}({1\over 2}({i\pi\over 6} - z_{k}^{+}))\over
-e^{-\phi_{-}}\sinh^{2}({i\pi\over 12})
+ e^{\phi_{-}}\sinh^{2}({1\over 2}({i5\pi\over 6} + z_{k}^{+}))}
\right]  
\times \left[\phi_{-} \rightarrow \phi_{+} \right] \,,
\label{zplus}
\ee
and from (I 4.15) we obtain a similar result for $z_{k}^{-}$.  In view
of the massless scaling limit (\ref{SYLlimit}), we set 
\be
\theta_{j} = \hat \theta_{j} - \ln \frac{n}{2} \,, \quad
z_{k}^{\pm} = \hat z_{k}^{\pm} - \ln \frac{n}{2} \,, \quad 
{i \pi\over 6}(b_{\pm} + 6) = \theta_{B}^{\pm} - \ln \frac{n}{2} \,, 
\quad n \rightarrow 0 \,,
\label{SYLlimit2}
\ee
and we obtain
\be
\prod_{j=1}^{N}
{\tanh({1\over 2}(\hat z_{k}^{+} -\hat \theta_{j}) )\over
\tanh({1\over 2}(\hat z_{k}^{+} -\hat\theta_{j}) + {i \pi\over 3})}
= 
{\cosh({1\over 2}(\hat z_{k}^{+} -\hat \theta_{B}^{-})-{i\pi\over 12})
\cosh({1\over 2}(\hat z_{k}^{+} -\hat \theta_{B}^{+})-{i\pi\over 12})
\over
\cosh({1\over 2}(\hat z_{k}^{+} -\hat \theta_{B}^{-})+{i5\pi\over 12})
\cosh({1\over 2}(\hat z_{k}^{+} -\hat \theta_{B}^{+})+{i5\pi\over 12})}
\,.
\label{zpluslim}
\ee
Finally, setting $\hat z_{k}^{+} = \hat x_{k} - {i \pi\over 3}$, we
obtain the Bethe Ansatz equations for $\hat x_{k}$ (cf. Eq.  (I 4.19)),
\be
\prod_{j=1}^{N}
{\tanh({1\over 2}(\hat x_{k} -\hat\theta_{j}) -{i \pi\over 6})\over
 \tanh({1\over 2}(\hat x_{k} -\hat\theta_{j}) +{i \pi\over 6})}
{\cosh({1\over 2}(\hat x_{k} -\hat \theta_{B}^{-})+{i\pi\over 4})
 \cosh({1\over 2}(\hat x_{k} -\hat \theta_{B}^{+})+{i\pi\over 4})
\over
 \cosh({1\over 2}(\hat x_{k} -\hat \theta_{B}^{-})-{i\pi\over 4})
 \cosh({1\over 2}(\hat x_{k} -\hat \theta_{B}^{+})-{i\pi\over 4})}
 = 1 \,,  \non  \\
k = 0 \,, 1 \,, \ldots \,, N \,. 
\label{realBAE}
\ee
The transfer matrix eigenvalues 
$\LLL (\theta | \theta_{1} \,, \ldots \,, \theta_{N})$ can be 
deduced from Eqs. (I 4.12), (I 4.17), (I 4.24). In the scaling 
limit (\ref{SYLlimit2}) (with $\theta = \hat \theta - \ln 
\frac{n}{2}$), we obtain
\be
\LLL \propto \Z(\hat \theta - \theta_{B}^{+}) 
\Z(\hat \theta - \theta_{B}^{-}) 
e^{\hat x_{0} -{1\over 2}(\theta_{B}^{+}+\theta_{B}^{-})}
\prod_{k=1}^{N} {Z(\hat \theta - \hat \theta_{k}) 
e^{\hat x_{k} - \hat \theta_{k}}\over 
{1\over 2}\sinh(\hat \theta - \hat \theta_{k})}
\prod_{k=0}^{N} 
\lambda_{\epsilon_{k}}(\hat \theta - \hat \theta_{k}) \,,
\label{eigenvalues}
\ee
where $Z(\theta)$ and $\Z(\theta)$ are 
given by (\ref{bulkZ}) and (\ref{masslessZfactor}), respectively;
\be
\lambda_{\epsilon}(\theta) = 
\sinh({\theta\over 2} + {\epsilon i\pi\over 6})
\cosh({\theta\over 2} - {\epsilon i\pi\over 6}) \,,
\ee
$\epsilon_{k} = \pm 1$ (see Eq. (I 4.20)), and 
$\hat x_{k}$ satisfy (\ref{realBAE}).

We introduce the densities $P_{\pm}(\hat \theta)$ of ``magnons'',
i.e., of real Bethe Ansatz roots $\{ \hat x_{k} \}$ with $\epsilon_{k}
= \pm 1$, respectively; and also the densities $\rho_{1}(\hat \theta)$
and $\tilde\rho(\hat \theta)$ of particles $\{ \hat \theta_{k} \}$ and
holes, respectively. The Bethe Ansatz equations 
(\ref{realBAE}) imply \footnote{The counting function should be
monotonic increasing, in order that the corresponding density (defined
as the derivative of the counting function) be nonnegative.}
\be
P_{+}(\hat \theta) + P_{-}(\hat \theta) = 
{1\over 2\pi} \left( \rho_{1} * \Phi \right)(\hat \theta) 
+ {1\over 2 \pi L}\left[
\Psi(\hat \theta - \theta_{B}^{+}) +
\Psi(\hat \theta - \theta_{B}^{-}) \right] \,,
\label{constraint1}
\ee
where
\be 
\Phi(\theta) &=& {1\over i} {\partial\over \partial \theta} \ln \left(
{\tanh({\theta\over 2} - {i\pi\over 6})\over
 \tanh({\theta\over 2} + {i\pi\over 6})}\right) =
 {4 \cosh\theta \sin {\pi\over 3}\over
 \cosh 2 \theta - \cos {2\pi\over 3}} = - \Phi_{YL}(\theta) \,, \non  \\
\Psi(\theta) &=& {1\over i} {\partial\over \partial \theta} \ln \left(
{\cosh({\theta\over 2} + {i \pi\over 4}) \over
 \cosh({\theta\over 2} - {i \pi\over 4})} \right) 
= {1\over \cosh \theta} \,,
\label{kernels}
\ee
and we have defined $\rho_{1}(\hat \theta)$ for negative values of
$\hat \theta$ to be equal to $\rho_{1}(|\hat \theta|)$.

The Yang equations (I 5.7) and the expression (\ref{eigenvalues}) for
the eigenvalues imply
\be
\rho_{1}(\hat \theta) + \tilde\rho(\hat \theta) &=& 
{\mu\over \pi} e^{\hat \theta } 
+ {1\over 2 \pi} \left( \rho_{1} * \Phi_{Z}\right) (\hat \theta) 
+ {1\over 2 \pi} \left( P_{+} * \Phi_{+} \right) (\hat \theta)
+ {1\over 2 \pi} \left( P_{-} * \Phi_{-} \right) (\hat \theta) \non \\
&+& {1\over 2 \pi L}\left[
{\partial\over \partial \hat \theta} 
Im \ln \Z(\hat \theta - \theta_{B}^{+})
+ {\partial\over \partial \hat \theta} 
Im \ln \Z(\hat \theta - \theta_{B}^{-})
\right] \,,
\ee
where
\be
\Phi_{Z}(\theta) = {\partial\over \partial \theta} Im \ln Z(\theta) 
\,, \qquad 
\Phi_{\pm}(\theta) = {\partial\over \partial \theta} Im \ln 
\lambda_{\pm}(\theta) \,,
\ee
and we have defined $P_{\pm}(\hat \theta)$ for negative values of
$\hat \theta$ to be equal to $P_{\mp}(|\hat \theta|)$.
Using the fact $\Phi_{\pm}(\theta) = \mp {1\over 2}\Phi(\theta)$, and 
using (\ref{constraint1}) to eliminate $P_{+}$, we obtain
\be
\rho_{1}(\hat \theta) + \tilde\rho(\hat \theta) &=& 
{\mu\over \pi} e^{\hat \theta} 
+ {1\over 2\pi}  \left( P_{-} * \Phi \right) (\hat \theta)
+ {1\over 2 \pi} \left( \rho_{1} * \left(\Phi_{Z} 
- {1\over 4 \pi} \Phi * \Phi \right)  \right) (\hat \theta) \non \\
&+& {1\over 2 \pi L}\Bigg[ 
{\partial\over \partial \hat \theta} 
Im \ln \Z (\hat \theta - \theta_{B}^{+})
- {1\over 4 \pi} \left( \Psi * \Phi \right)(\hat \theta - \theta_{B}^{+})
\non \\
&+& 
{\partial\over \partial \hat \theta} 
Im \ln \Z (\hat \theta - \theta_{B}^{-})
- {1\over 4 \pi} \left( \Psi * \Phi \right)(\hat \theta - \theta_{B}^{-})
\Bigg] \,. 
\ee
With the help of the identities
\be
\Phi_{Z}(\theta)  
- {1\over 4 \pi} \left( \Phi * \Phi \right)(\theta)  &=& -\Phi(\theta)  
\,, \non  \\
{\partial\over \partial \theta} Im \ln \Z (\theta)
- {1\over 4 \pi} \left( \Psi * \Phi  \right)(\theta ) &=& 0 \,,
\ee 
we obtain the simple result
\be
\rho_{1}(\hat \theta) + \tilde\rho(\hat \theta) &=& 
{\mu\over \pi} e^{\hat  \theta} 
+ {1\over 2\pi}  \left( P_{-} * \Phi \right)(\hat \theta)
- {1\over 2\pi}  \left(\rho_{1} * \Phi \right)(\hat \theta)
\,.
\label{constraint2}
\ee

Proceeding as in I, we obtain the TBA equations \footnote{This set of TBA 
equations is the same as for the case of periodic boundary conditions, which 
was first conjectured in \cite{RTV} (see also \cite{DDT}) and later derived 
from the SYL $S$ matrix in \cite{Ah1} and generalized in \cite{MS1}.}
\be
r e^{\hat \theta} &=& \epsilon_{1}(\hat \theta) 
- {1\over 2 \pi} \left( \Phi *(L_{1} -  L_{2}) \right)(\hat \theta) 
\,, \non  \\
0 &=& \epsilon_{2}(\hat \theta) 
+ {1\over 2 \pi} \left( \Phi * L_{1} \right)(\hat \theta) \,, 
\label{TBA}
\ee
where 
\be
L_{i}(\hat \theta) &=& \ln \left( 1 + e^{-\epsilon_{i}(\hat \theta)} 
\right) \,, \qquad r = \mu R \,, \non \\
\epsilon_{1} &=& \ln \left( {\tilde \rho\over \rho_{1}} \right) \,, 
\qquad 
\epsilon_{2} = \ln \left( {P_{+}\over P_{-}} \right) \,.
\ee
Moreover, the boundary entropy of one boundary is given (up to an additive 
constant) by 
\be
\ln g = {2\over 4\pi}\int_{-\infty}^{\infty}d \hat \theta\ 
\Psi(\hat \theta - \theta_{B})\  L_{2}(\hat \theta) 
\,,
\label{beresult}
\ee
where we have included the factor 2 in order to account for 
contributions from both right-movers and left-movers.
In the UV limit $\theta_{B} \rightarrow -\infty$, 
the integrand is nonvanishing for $\hat \theta \rightarrow -\infty$;
similarly, the IR limit $\theta_{B} \rightarrow \infty$ requires
$\hat \theta \rightarrow \infty$. Using the results 
$L_{2}(-\infty) = \ln \left(2+\sqrt{2}\right)$, 
$L_{2}(\infty) =  \ln 2$ which follow from the 
TBA Eqs. (\ref{TBA}), we obtain
\be
\ln {g^{UV}\over g^{IR}} = {1\over 2}\ln \left(
{1+\sqrt{2}\over \sqrt{2}}\right)
\,.
\label{SYLflow1}
\ee
This is precisely the ratio of $g$ factors corresponding to the
${\cal M}(3/8)$ conformal boundary conditions $(1 \,, 3)$ and $(1 \,, 2)$ 
\be
\ln {g_{(1 \,, 3)}\over g_{(1 \,, 2)}} = {1\over 2}\ln \left(
{1+\sqrt{2}\over \sqrt{2}}\right)
\,,
\label{SYLflow2}
\ee
as one can verify from Eq. (\ref{SYLgfactors}).  
Hence, the proposed boundary $S$ matrix is consistent with massless
flow away from the UV conformal boundary condition; namely, from the
CBC $(1 \,, 3)$ to the CBC $(1\,, 2)$.  In the ${\cal SM}(2/8)$
description, this corresponds to the flow from the SCBC $(1 \,, 3)$ to
the SCBC $(1\,, 4)$.  A plot of $\ln g$ as a function of $\theta_B$ is
given in Fig.  1.  For convenience, a constant has been added so that
the UV value is $\frac{1}{2} \ln (1+ \sqrt{2})$ and the IR value is
$\frac{1}{2} \ln \sqrt{2}$.
\begin{figure}[htb]
	\centering
	\epsfxsize=0.7\textwidth\epsfbox{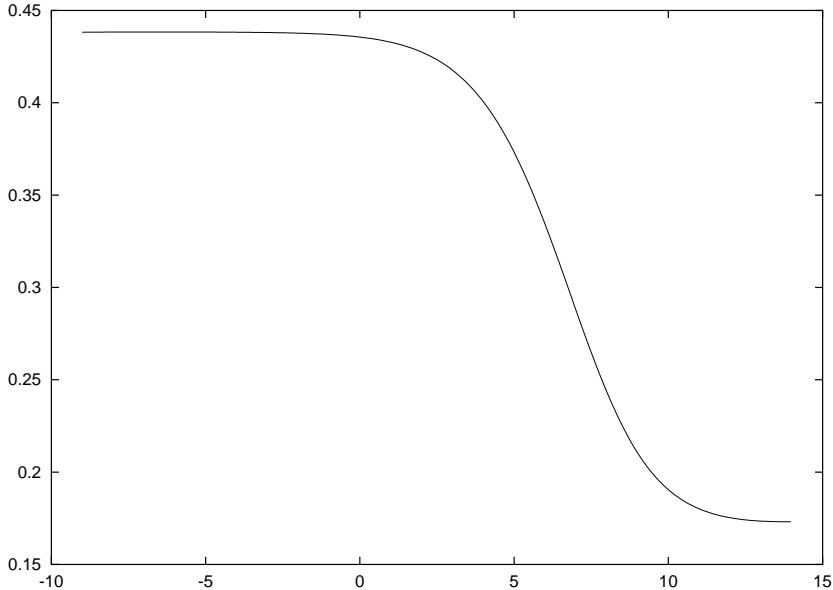}
	\caption[xxx]{\parbox[t]{.4\textwidth}{
	Boundary entropy: $\ln g$ vs. $\theta_B$.}
	}
	\label{figclosed}
\end{figure}

\section{Discussion}\label{sec:discuss}

We have proposed the boundary $S$ matrix 
(\ref{boundarySmatrix}) - (\ref{zeta}), (\ref{desired}), 
(\ref{CDDsolution}) for the boundary SYL model defined by the action
(\ref{boundSYL}). Some support for this conjecture is provided by the 
fractional-spin integral of motion (\ref{boundSUSY}), 
(\ref{boundSUSY2}), and by the massless boundary flow 
(\ref{SYLflow1}), (\ref{SYLflow2}).
Several important problems remain to be solved, including
the relation of the parameter $\lambda_{B}$ in the action to the 
parameter $b$ of the boundary $S$ matrix; and the identification of 
the operator $\Theta$ in (\ref{boundSUSY}) with the operator 
$\gamma (-1)^{F}$ in (\ref{boundSUSY2}). It would also be interesting to 
consider other conformal boundary conditions, as well as
extend the present study to the full family of integrable models with 
$N=1$ supersymmetry \cite{Sc}, \cite{MS2}.

\section*{Acknowledgments}

We thank O. Alvarez, P. Dorey, D. Kastor, P. Pearce, F. Ravanini, R.
Tateo and M. Walton for helpful discussions and/or correspondence. 
C.A. thanks Universit\`a di Bologna and University of Miami for
hospitality where part of this work was done.  This work was supported
in part by KOSEF 1999-2-112-001-5 (C.A.) and by the National Science
Foundation under Grant PHY-9870101 (R.N.).

\appendix

\section{SYL model as restriction of ZMS model}

Here we show that the scaling supersymmetric Yang-Lee model is a
restriction of the Zhiber-Mikhailov-Shabat model \cite{DB}, \cite{ZMS},
whose action is given by
\be
A = \int d^{2}x\ (\partial_{\mu}\varphi)^{2} + {m^{2}\over \gamma^{2}}
\left( e^{i\sqrt{8 \gamma}\varphi} + e^{-i\sqrt{2 \gamma}\varphi} \right) 
\,. \label{ZMS}
\ee 
This is the $A^{(2)}_{2}$ imaginary coupling affine Toda field 
theory, whose $S$ matrix was found by Izergin and Korepin \cite{IK}.
We follow closely the paper \cite{Ta} of Tak\'acs, to
which we shall refer as II. 

It is useful to first recall the related work \cite{Sm2}
of Smirnov. There it is observed that, for 
\be
\gamma = {\pi r\over s} \,,
\ee
the ZMS model is the $(1 \,, 2)$ perturbation of the minimal model
${\cal M}(r/s)$.  Indeed, one can regard the first two terms in
the action (\ref{ZMS}) as the action for ${\cal M}(r/s)$, and the
third term as the $(1 \,, 2)$ perturbation.  The $S$ matrix of the
perturbed model can be obtained as the RSOS restriction of the
$A^{(2)}_{2}$ $S$ matrix, using the model's $U_{q}(sl(2))$ symmetry,
where $q=e^{i \pi^{2}/\gamma}$.

In II, it is observed that, for 
\be
\gamma = 4 \gamma' = {4\pi r'\over s'} \,,
\ee
the ZMS model is the $(1 \,, 5)$ perturbation of the minimal model
${\cal M}(r'/s')$.  Indeed, one can regard the first and third terms
in the action (\ref{ZMS}) as the action for ${\cal M}(r'/s')$, and the
second term as the $(1 \,, 5)$ perturbation.  The $S$ matrix of the
perturbed model can be obtained as the RSOS restriction of the
$A^{(2)}_{2}$ $S$ matrix, using the model's $U_{q'}(sl(2))$ symmetry,
where $q'=e^{i \pi^{2}/\gamma'}=q^{4}$.

We have suggested in Section \ref{sec:bulkSYL} that the SYL model can
be regarded as the $(1 \,, 5)$ perturbation of ${\cal M}(3 /8)$. 
We now proceed to compute the latter's $S$ matrix following II, and we
shall find that it coincides (up to a scalar factor) with Eq.
(\ref{bulkRmatrix}). For ${\cal M}(3/8)$ we have $r'=3$, $s'=8$; 
hence, $q'=q=e^{2 i \pi/3}$.
The first positive integer $p$ for which $q'^{p} = \pm 1$ is $p=3$. 
Hence, the maximum spin is $j_{max}= {p\over 2} -1 = {1\over 2}$. 
Thus, the model contains ``charged'' kinks 
$K_{0\ {1\over 2}} = K_{{1\over 2}\ 0}$ which we denote by $c$, and 
``neutral'' kinks $K_{0\ 0} = K_{{1\over 2}\ {1\over 2}}$ which we 
denote by $n$. Since (II 23)
\be 
\xi = {2\over 3} \left( {\pi \gamma \over 2\pi - \gamma} \right) = 
2\pi \,,
\ee
the model contains neither breathers nor higher kinks. The $S$ matrix 
is expressed in terms of the rapidity variable $y = e^{\pi 
\theta/\xi} = e^{\theta/2}$. The $c\ c \rightarrow c\ c$ amplitude
is given by (II 43) - (II 45) 
\setlength{\unitlength}{0.012500in}
\be
\begin{picture}(370,45)(20,780)
\thinlines
\put( 20,780){\line( 1, 1){ 40}}
\put( 20,795){\makebox(0,0)[lb]{0}}
\put( 20,820){\line( 1,-1){ 40}}
\put( 55,795){\makebox(0,0)[lb]{0}}
\put( 75,795){\makebox(0,0)[lb]{\smash{= 
${\displaystyle \frac{y^2}{q}-\frac{q}{y^2}-\frac{1}{q}+q +
\frac{y^2}{q^5}-\frac{q^5}{y^2}-\frac{1}{q}+q 
= 2 i \sqrt{3}- 2 \sinh \theta} \,. $ }}}
\put( 35,780){\makebox(0,0)[lb]{$\frac{1}{2}$}}
\put( 35,810){\makebox(0,0)[lb]{$\frac{1}{2}$}}
\end{picture} 
\ee 
The $n\ n \rightarrow n\ n$ amplitude
is given by (II 46), (II 40)
\setlength{\unitlength}{0.012500in}
\be
\begin{picture}(370,45)(20,780)
\thinlines
\put( 20,780){\line( 1, 1){ 40}}
\put( 20,795){\makebox(0,0)[lb]{0}}
\put( 20,820){\line( 1,-1){ 40}}
\put( 55,795){\makebox(0,0)[lb]{0}}
\put( 75,795){\makebox(0,0)[lb]{\smash{= 
${\displaystyle \frac{q^6y^2+y^2q^8-q^8-q^4y^2+y^2-q^{10}y^2
+y^4q^2-y^2q^2}{y^2q^5}
= 2 i \sqrt{3}+ 2 \sinh \theta} \,. $ }}}
\put( 35,780){\makebox(0,0)[lb]{0}}
\put( 35,810){\makebox(0,0)[lb]{0}}
\end{picture}  
\ee 
The $c\ c \rightarrow n\ n$ and $n\ n \rightarrow c\ c$ amplitudes are
equal, are are given by (II 48)
\setlength{\unitlength}{0.012500in}
\be
\begin{picture}(370,45)(20,780)
\thinlines
\put( 20,780){\line( 1, 1){ 40}}
\put( 20,795){\makebox(0,0)[lb]{0}}
\put( 20,820){\line( 1,-1){ 40}}
\put( 55,795){\makebox(0,0)[lb]{0}}
\put( 75,795){\makebox(0,0)[lb]{\smash{= 
${\displaystyle i\frac{(q^4-1)(y^2-1)}{q^2y}
= 2 \sqrt{3} \sinh \frac{\theta}{2}} \,. $ }}}
\put( 35,780){\makebox(0,0)[lb]{$\frac{1}{2}$}}
\put( 35,810){\makebox(0,0)[lb]{0}}
\end{picture} 
\ee 
Finally, the $n\ c$ forward scattering 
and reflection amplitudes are given by (II 46), (II 40)
\setlength{\unitlength}{0.012500in}
\be
\begin{picture}(370,45)(20,780)
\thinlines
\put( 20,780){\line( 1, 1){ 40}}
\put( 20,795){\makebox(0,0)[lb]{0}}
\put( 20,820){\line( 1,-1){ 40}}
\put( 55,795){\makebox(0,0)[lb]{$\frac{1}{2}$}}
\put( 75,795){\makebox(0,0)[lb]{\smash{= 
${\displaystyle \frac{(y^2+q^6)(y^2-1)}{y^2q^3}
= 2 \sinh \theta}$ }}}
\put( 35,780){\makebox(0,0)[lb]{0}}
\put( 35,810){\makebox(0,0)[lb]{$\frac{1}{2}$}}
\end{picture} 
\ee 
and
\setlength{\unitlength}{0.012500in}
\be
\begin{picture}(370,45)(20,780)
\thinlines
\put( 20,780){\line( 1, 1){ 40}}
\put( 20,795){\makebox(0,0)[lb]{$\frac{1}{2}$}}
\put( 20,820){\line( 1,-1){ 40}}
\put( 55,795){\makebox(0,0)[lb]{0}}
\put( 75,795){\makebox(0,0)[lb]{\smash{= 
${\displaystyle -\frac{(q^4-1)(y^2+q^6)}{yq^5}
= 2i \sqrt{3} \cosh \frac{\theta}{2}} \,, $ }}}
\put( 35,780){\makebox(0,0)[lb]{$\frac{1}{2}$}}
\put( 35,810){\makebox(0,0)[lb]{$\frac{1}{2}$}}
\end{picture} 
\ee 
respectively.  Identifying $n$ and $c$ as the Boson and Fermion
(respectively) of the SYL model, we see that the above amplitudes
coincide with those in $2 (\sinh \theta) R(\theta)$, where $R(\theta)$
is the matrix (\ref{bulkRmatrix}).  That is, the SYL model is indeed a
restriction of the ZMS model, corresponding to the $(1 \,, 5)$
perturbation of ${\cal M}(3/8)$.

\section{Solution of constraint Eq. (\ref{constraint})}

Here we solve Eq. (\ref{constraint}), which for simplicity we 
now write as
\be
\left( {\cal A}_{+}{\cal A}_{-} - {\cal B}^{2} \right)
\Big\vert_{\theta = iv} = 0 \,.
\label{newform}
\ee 
Using the definitions of ${\cal A}_{+}$, ${\cal A}_{-}$, and ${\cal B}$ 
given in (\ref{boundelem1}), (\ref{boundelem2}), and introducing the 
variable $t \equiv \sin^{2} {v\over 2}$, Eq. (\ref{newform}) can be 
brought to the form
\be
\left( t-{1\over 2}\right)\Bigg[ 
t^{2} + t \left(-1 + {\sqrt{3}\over 2} + 
e^{-2\phi}\left({3\over 4} -  {\sqrt{3}\over 2}\right) \right) \non  \\
+ {7\over 16} - {\sqrt{3}\over 4} 
+e^{-2\phi}\left({\sqrt{3}\over 2} - {7\over 8} \right) 
+e^{-4\phi}\left({7\over 16} -  {\sqrt{3}\over 4}\right) 
\Bigg] = 0 \,.
\ee
We discard the solution $t={1\over 2}$, which corresponds to a 
fixed value of $v$ (and hence, $b$). The two remaining solutions are
$t = {1\over 4}(\gamma \mp \sqrt{\Delta})$, where
\be
\gamma = 2 - \sqrt{3} 
+ e^{-2\phi}\left( \sqrt{3} - {3\over 2} \right) \,, \quad 
\Delta = e^{-2\phi}\left( 2 - \sqrt{3} \right) 
+ e^{-4\phi}\left( \sqrt{3} - {7\over 4} \right) \,.
\ee 
In terms of the parameter $\zeta$ defined by
\be 
\cos \zeta = 1 - e^{-2 \phi}\left(1 - {\sqrt{3}\over 2}\right) \,,
\ee
we have
\be
\gamma = 2 - \sqrt{3} \cos \zeta \,, \qquad \Delta = \sin \zeta
\,;
\ee 
and therefore, 
\be
t = {1\over 2}\left[ 1 - \cos \left( \zeta \mp {\pi\over 6} \right) \right] \,.
\ee
Finally, recalling the definition $t = \sin^{2} {v\over 2}$, we 
arrive at the remarkably simple result
\be 
\zeta = v \pm {\pi\over 6} \,,
\ee 
which is quoted in text (\ref{zetarelation}).

\end{document}